\documentstyle[pra,aps]{revtex}
\begin{document}
\draft
\title{ A note on the supplementary variables in spin-measuring equipments 
      in the EPR-Bell experiment }
\author{ Won Young Hwang \thanks{e-mail: hwang@chep6.kaist.ac.kr}
  and In Gyu Koh  }
\address{  Department of Physics,  KAIST, Yusung, Taejon, Korea }
\author{ Yeong Deok Han \thanks{e-mail: ydhan@core.woosuk.ac.kr}}
\address{ Department of Physics, Woosuk University, 
      Hujeong, Samrye, Wanju, Cheonbuk, Korea } 
\maketitle
\begin{abstract}
We discuss supplementary ( or hidden ) variables in spin-measuring
equipments in EPR-Bell experiment. This theme was considered in a Bell's
later work. We generalize it. First, we show why the original
supplementary variable $\lambda$ is not to be regarded to include 
supplementary variables in spin-measuring equipments ( why supplementary
variables should be introduced additionally in spin-measuring equipments )
Next, we show the followings. When the supplementary variables 
introduced in spin-measuring equipments have local correlations, the
Bell inequality is recovered. On the other hand, when they have nonlocal
correlations, the Bell inequality is not recovered. This fact is in 
accord with the fact that the Bell inequality is derived for local 
realistic models.

\end{abstract}

\section{Introduction}
Since the discovery of quantum mechanics, there have been many 
controversies$^{(1)}$  concerning whether 
quantum mechanics is compatible with realism. In the Copenhagen 
interpretation$^{(1,2)}$  of quantum mechanics,
it was advocated that the
world is no more compatible with realism and thus it is useless
to make efforts to find pictures of what are happening in the
world. The Copenhagen interpretation has been accepted by most 
physicists. However, some realistic interpretations ( or realistic
models )
of quantum mechanics have been discovered$^{(3,4,5)}$. 
 These realistic interpretations have nonlocal
 characters, which seem to be awkward to physicists acquainted
with local realism that has been very successful for a long time 
in natural sciences. Thus it is an important problem whether
{\it local} realistic interpretation of quantum mechanics is also
possible
( in other words whether quantum mechanical predictions can be 
reproduced by local realistic models ).
Bell$^{(6,9)}$  gave a surprising answer to this question;
the local realism cannot coexist with quantum mechanics ( The Bell
theorem ). This was shown by the fact that an inequality ( the Bell
inequality ) that must be satisfied by all local realistic models 
is violated by quantum mechanics. 
In order to know which one is right, experiments have to be done.
In spite of real experiments done$^{(10)}$,
there still remain some controversies concerning the interpretations of 
the results, mainly concerning the detection-loophole$^{(9,11)}$.

By the way, in ( local or nonlocal ) realistic models for quantum
mechanics, supplementary ( or hidden ) variable $\lambda$ should
be introduced to systems in order to explain intrinsic statistical
character of quantum mechanics. 
In Bell's original paper$^{(6)}$, the supplementary variable 
$\lambda$ corresponds to only the state of particle pairs from source
of the EPR ( Einstein-Podolsky-Rosen )-Bell experiment$^{(9,12)}$. 
We will show this fact in section 3. In a Bell's later
work$^{(7)}$, the conditions are generalized: the supplementary
variables are introduced also to the spin-measuring equipments of 
the EPR-Bell experiment and deterministic one is relaxed to probabilistic 
one. Bell showed$^{(7)}$  that Bell inequality is recovered
even in those cases. The purpose of this note is to make the recovering
process more general and explicit. 
\section{ the Bell inequality}
Bell$^{(6,9)}$  showed that it is impossible for
 any local realistic models to reproduce
all the predictions of quantum mechanics for EPR experiment$^{(9,12)}$.
The experiment is described as follows. 
A source emits pairs of spin-$\frac{1}{2}$ particles, in the singlet state
$\frac{1}{\sqrt{2}}(|z+\rangle|z-\rangle - |z-\rangle|z+\rangle)$.
 After particles have
separated, one performs correlated measurements of their spin components
along arbitrary directions $a$ and $b$ ( Fig.1 ). Each measurement
can yield two results, $\pm1$. 
For the singlet state, quantum mechanics predicts some correlations
between such measurements on the two particles.
We denote by $p_{\pm\pm}(a,b)$ the probabilities of obtaining 
the result $\pm1$ along $a$ at 
one place A and $\pm1$ along $b$ at the other place B. The quantity
 \begin{equation}
  E(a,b) = p_{++}(a,b) + p_{--}(a,b)
                   -p_{+-}(a,b) - p_{-+}(a,b)
  \label{1}
 \end{equation}
is the correlation function of measurements of two particles.

In ( deterministic ) realistic models,
it is assumed that the outcome of measurements of spin is determined
by some supplementary ( or hidden ) variable $\lambda$ 
and by the directions $a$ and $b$  of spin-measuring equipments.
 The supplementary variable $\lambda$  
is a random variable; the probability distribution of which is
given by some function
$\rho_{ab}(\lambda)$ having the normalization property
 \begin{equation}
  \int \rho_{ab}(\lambda)\hspace{0.14cm} d\lambda = 1,
  \label{2}
 \end{equation}
and depending on $a$,$b$ and on the state $\psi$ of the pair of particles. 
The outcomes of spin-measurements are assumed to be determined
by $\lambda$ and by direction of each corresponding spin-measurement;
 \begin{eqnarray}
  S_A(a)= f(a,\lambda),\hspace{0.5cm}
  S_B(b) &=& g(b,\lambda). \hspace{1cm} \mbox{the locality condition}
  \label{3}
 \end{eqnarray}
The functions $f$ and $g$ depend on the particular properties
of measuring 
equipments and the range of which are $\{ \pm 1 \} $.

The combination of correlation functions
 \begin{equation}
  S=E(a,b)+E(a,b^\prime)+E(a^\prime,b)    
   -E(a^\prime,b^\prime)
  \label{4}
 \end{equation}
is constrained by these properties of local realistic models.
$E(a,b)$ in the case of local realistic models is given 
 by Eq.(\ref{1}) and Eq.(\ref{3});
  \begin{eqnarray}
  E(a,b)
 &=&\int f(a,\lambda)g(b,\lambda) \rho_{ab}(\lambda)d\lambda
  \label{5}
  \end{eqnarray} 
With these equations,
           \begin{eqnarray}
 \mid S \mid
&=&\mid E(a,b)+E(a,b^\prime)+E(a^\prime,b)
       -E(a^\prime,b^\prime)\mid,\nonumber\\
&=&|\int f(a,\lambda)g(b,\lambda) \rho_{ab}(\lambda)d\lambda
+\int
f(a,\lambda)g(b^\prime,\lambda)
                 \rho_{a^\prime,b}(\lambda)d\lambda\nonumber\\
& & +\int f(a^\prime,\lambda)g(b,\lambda)
                 \rho_{a,b^\prime}(\lambda)d\lambda
  -\int f(a^\prime,\lambda)g(b^\prime,\lambda)
                      \rho_{a^\prime,b^\prime}(\lambda)d\lambda|.
                                                      \label{6}
\end{eqnarray}
If $\lambda$ corresponds to only particle pairs from source, it can be 
justified from locality condition that the probability distribution
$\rho_{ab}(\lambda)$ is independent upon $a$ and $b$, the directions
of spin-component being measured at $A$ and $B$ respectively,
\begin{equation}
  \rho_{ab}(\lambda)=\rho_{a,b^\prime}(\lambda)
 =\rho_{a^\prime,b}(\lambda)=\rho_{a^\prime,b^\prime}(\lambda)
 \equiv \rho(\lambda).
 \label{7}
\end{equation}
Thus,
\begin{eqnarray}
  |S|&=&\mid \int [f(a,\lambda)\{g(b,\lambda)+g(b^\prime,\lambda)\}
      +f(a^\prime,\lambda)\{g(b,\lambda)
      -g(b^\prime,\lambda)\}]\rho(\lambda)d\lambda \mid, \nonumber \\
  &\leq& \int\{ |f(a,\lambda)| |g(b,\lambda)+g(b^\prime,\lambda)|
             + |f(a^\prime,\lambda)| |g(b,\lambda)-g(b^\prime,\lambda)|\}
               \rho(\lambda)d\lambda, \nonumber\\
  &\leq& \int\{ |g(b,\lambda)+g(b^\prime,\lambda)|
             + |g(b,\lambda)-g(b^\prime,\lambda)|\}
               \rho(\lambda)d\lambda, \nonumber\\
  &\leq& \int 2 \rho(\lambda) d\lambda = 2.
  \label{8}
\end{eqnarray}
Eq.(\ref{8}) is derived from the facts that $|f|\leq1$ and $|g|\leq1$.
This is the Bell inequality,
 \begin{equation}
   \mid E(a,b)+E(a,b^\prime)+E(a^\prime,b)
       -E(a^\prime,b^\prime)\mid \leq 2,
  \label{9}
 \end{equation}
violated by quantum mechanics at certain cases.

We might think over broader classes of local realistic
models where the assumption (\ref{3}) is a little relaxed.
These are 'the stochastic local realistic model$^{(7,13,14)}$ '
and 'the contextual local realistic model$^{(9)}$'.
In the former case, determinism is relaxed;
the probability that spin-measurements along a direction $a$
( $b$ ) at A ( at B ) for a certain $\lambda$ will give an outcome
$m$ ( $n$ ) ( $m,n= \pm 1$ ) is given $p(m|a,\lambda)$  
 ( $p(n|b,\lambda)$ ).
In this case, after some algebraic manipulations we obtain,
for example,
\begin{eqnarray}
 E(a,b)&=&\int \bar{f}(a,\lambda) \bar{g}(b,\lambda)
	     \rho_{ab}(\lambda) d\lambda, \nonumber\\
\mbox{where}\hspace{0.5cm}& &
	     \bar{f}(a,\lambda)=\sum_m m p(m|a,\lambda),\hspace{0.5cm}
             \bar{g}(b,\lambda)=\sum_n n p(n|b,\lambda).
             \label{10}
\end{eqnarray}
,which is the same form as that of Eq.(\ref{5}). Thus, by simply replacing 
$f(a,\lambda)$ and $g(b,\lambda)$ by $\bar{f}(a,\lambda)$
and $\bar{g}(b,\lambda)$, we can generalize all proofs in this
letter to the case of stochastic local realistic models.

In the latter case, contextual interactions within light-cone 
is permitted; outcomes of spin-measurements depend also on
direction of spin-measurements on the other side, if the two 
events are not space-likely separated,
 \begin{equation}
  S_A(a)= f(a,b,\lambda),\hspace{0.5cm}
  S_B(b)= g(a,b,\lambda). 
  \label{11}
 \end{equation}
In this case, to recover the Bell inequality, Eq.(\ref{11}) is
reduced to Eq.(\ref{3}) by space-likely separating
the events of detections of spin at A from those at B. 

\section{ Supplementary ( or hidden ) variables in spin measuring 
         equipments }
If spin-measuring equipments are composed of similar ones as the 
detected particles, it is natural to introduce additional supplementary
variables $\lambda_i$  ( $i=a,a^\prime,b,b^\prime$ ) ( Fig.1 ).
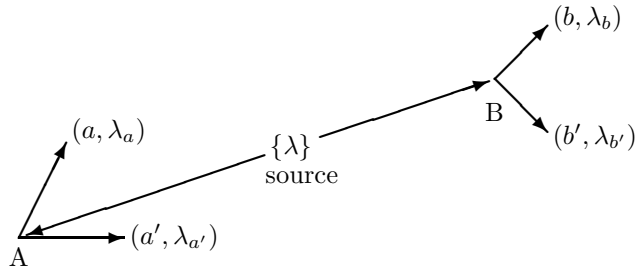
\begin{figure}[p]
    \begin{picture}(200,100)(-140,10)
          \thicklines
          \put(0,0){\vector(1,2){18}}
          \put(0,0){\vector(1,0){40}}
          \put(180,60){\vector(1, 1){20}}
          \put(180,60){\vector(1,-1){20}}
          
          \put(20,40){\makebox(0,0)[l]{($a       ,\lambda_a$)}}   
          \put(42,0 ){\makebox(0,0)[l]{($a^\prime,\lambda_{a^\prime}$)}}
          \put(202,83){\makebox(0,0)[l]{($b      ,\lambda_b$)}}
         \put(202,38){\makebox(0,0)[l]{($b^\prime,\lambda_{b^\prime}$)}}
 
          \put(93,31 ){\vector(-3,-1){90}}
          \put(114,38){\vector(3 , 1){64}}

          \put( 95,34){\makebox(0,0)[l]{$\{ \lambda \}$}}
         %\put( 79,58){\makebox(0,0)[l]{$\{ \lambda^\prime \}$}}

          \put(0  ,-4){\makebox(0,0)[t]{A}}
          \put(180,51){\makebox(0,0)[t]{B}}               
          \put(107,25 ){\makebox(0,0)[t]{source}}
         %\put(90,68 ){\makebox(0,0)[b]{hidden source}}               

         %\thinlines 
         %\put(78, 52){\vector(-3,-2){71}}
         %\put(102,60){\vector( 1, 0){71}}
         %\multiput(90,60)(-3,-2){30}{\line(-3,-2){1}}
    \end{picture}
    \vspace{1cm}
    \caption{ $\lambda_i$ ( $i=a,a^\prime,b,b^\prime$ ) are supplementary
  variables in spin-measuring equipments.}
\end{figure}
Then, since $\lambda_i$ as well as $\lambda$ take parts in measuring process,
the outcome of spin-measurements are determined by $\lambda_i$
as well as $\lambda$ and directions of spin-measuring equipments,
\begin{eqnarray}
     S_A(a)=f(a,\lambda,\lambda_a)\hspace{1cm} 
     S_B(a)=g(b,\lambda,\lambda_b)\nonumber\\
     S_A(a^\prime)=f(a^\prime,\lambda,\lambda_{a^\prime})\hspace{1cm} 
     S_B(b)=g(b^\prime,\lambda,\lambda_{b^\prime}) 
     \label{12}
\end{eqnarray}
 One may argue that we do not need to additionally introduce these variables
$\lambda_i$, because the original variable $\lambda$ can be regarded to
include all these variables, since there is no specification that 
$\lambda$ corresponds to only source particle pairs.  We may answer to this 
argument as follows. As shown in section 2 the independence of
$\rho_{ab}(\lambda)$ on a and b ( Eq.(\ref{7}) ) was used 
for derivation of Bell inequality. And this independence was justified
by locality only when $\lambda$ corresponds to only source particles
which are space-likely separated from the measuring process at each side.
 Thus, if $\lambda$ corresponds to spin-measuring equipments
as well as to source particles, then the independence condition 
( Eq.(\ref{7}) ) is not given.
So, Bell inequality is not obtained immediately without
some operations which will be described in the following.                    
The above demonstration is applied also to the slightly different 
version$^{(15,13,14)}$  of Bell inequality 
( where the locality condition ( Eq.(\ref{3})) is replaced by 
factorizability ( Eq.($2^\prime$) of Ref.(15) ), because
the independency condition ( Eq.(\ref{7})) is also assumed implicitly
 in Ref.(15).  

\section{Recovering the Bell inequality}
The probability distributions
$\rho_{pq}(\lambda,\lambda_p,\lambda_q)$  
 ( $ p=a,a^\prime \hspace{0.3cm} q=b,b^\prime$ ) of $\lambda$
 corresponding to particles pairs
from source and $\lambda_i$ ( $ i=a,a^\prime,b,b^\prime $ ) corresponding
to spin-measuring equipments are not independent upon $a$ and $b$.
With these general distributions
$\rho_{pq}(\lambda,\lambda_p,\lambda_q)$,
we obviously cannot recover Bell inequality. However, the locality 
will constraint the $\rho_{pq}(\lambda,\lambda_p,\lambda_q)$ in some
ways. The first possibility is that they factorize to each part, 
for example,
\begin{equation}
    \rho_{ab}(\lambda,\lambda_a,\lambda_b)=
    \rho_{ab}(\lambda)\rho_{ab}(\lambda_a)\rho_{ab}(\lambda_b).
    \label{13}
\end{equation}
This factorization is a physically plausible assumption, since this is
equivalent to assuming that source and measuring equipments of each 
side are independent upon each other.
By locality we have, for example,
\begin{equation}
           \rho_{ab}(\lambda)\rho_{ab}(\lambda_a)\rho_{ab}(\lambda_b)=
           \rho(\lambda)\rho_a(\lambda_a)\rho_b(\lambda_b).
   \label{14}
\end{equation}
Then we have, for example,
\begin{eqnarray}
 E(a,b)&=&\int\rho(\lambda)\rho_a(\lambda_a)\rho_b(\lambda_b)
 f(a,\lambda,\lambda_a)g(b,\lambda,\lambda_b)
                               d\lambda d\lambda_a d\lambda_b
   \nonumber\\
 &=&\int[\int f(a,\lambda,\lambda_a)\rho_a(\lambda_a) d\lambda_a]
        [\int g(a,\lambda,\lambda_b)\rho_b(\lambda_b) d\lambda_b]
        \rho(\lambda)d\lambda
   \nonumber\\
 &=&\int \bar{F}(a,\lambda)\bar{G}(b,\lambda)\rho(\lambda)d\lambda,
   \label{15} \\
  \mbox{where}\hspace{0.5cm}  \bar{F}(a,\lambda)&\equiv&
           \int f(a,\lambda,\lambda_a)\rho_a(\lambda_a) d\lambda_a ,
 \hspace{0.5cm}
                     \bar{G}(b,\lambda)\equiv
           \int g(a,\lambda,\lambda_b)\rho_b(\lambda_b) d\lambda_b 
          \nonumber
\end{eqnarray}
which is the same form as that of Eq.(\ref{5}) so that we can recover
Bell inequality.
Similarity between Eq.(\ref{10}) and Eq.(\ref{15}) 
can be used in order that deterministic local realistic models with 
some extra variables $\lambda_i$ in spin-measuring equipments
( Eq.(\ref{15}) ) emulate stochastic local realistic models
 ( Eq.(\ref{10}) ).

The next possibility we consider is the case where there exist a joint
probability 
distribution $\rho(\lambda,\lambda_a,\lambda_{a^\prime},\lambda_b,
\lambda_{b^\prime})$ which returns
$\rho_{pq}(\lambda,\lambda_p,\lambda_q)$ 
as marginal probability distributions, for example,
\begin{equation}
  \rho_{ab}(\lambda,\lambda_a,\lambda_b)
       = \int \rho(\lambda,\lambda_a,\lambda_{a^\prime},\lambda_b,
         \lambda_{b^\prime}) d\lambda_{a^\prime} d\lambda_{b^\prime}.
         \label{16}
\end{equation}
In this case we have, for example,
\begin{equation}
    E(a,b)=\int f(a,\lambda,\lambda_a)g(b,\lambda,\lambda_b)
   \rho(\lambda,\lambda_a,\lambda_{a^\prime},
   \lambda_b, \lambda_{b^\prime})
d\lambda d\lambda_a d\lambda_{a^\prime} d\lambda_b d\lambda_{b^\prime}.
    \label{17}  
\end{equation}
 We define,
\begin{equation}
 \lambda \otimes \lambda_a \otimes \lambda_{a^\prime} \otimes \lambda_b
 \otimes \lambda_{b^\prime} \equiv \tilde{\lambda}.
 \label{18}
\end{equation}
And we regard $f(a,\lambda,\lambda_a)$ ( $g(b,\lambda,\lambda_b)$ )
as a function of $ a$ and $ \tilde{\lambda} $
 ( $b$ and $ \tilde{\lambda}$ )
\begin{equation}
 f(a,\lambda,\lambda_a)\equiv f(a,\tilde{\lambda}),
 \hspace{1cm} g(b,\lambda,\lambda_b)\equiv g(b,\tilde{\lambda}).
 \label{19}
\end{equation}
Then we have, for example,
\begin{equation}
 E(a,b)= \int f(a,\tilde{\lambda}) g(b,\tilde{\lambda}) 
              \rho(\tilde{\lambda}) d\tilde{\lambda}    
              \label{20}
\end{equation}
which is the same form as that of Eq.(\ref{5}).
Therefore, we can recover Bell inequality. 

Let us now discuss about the physical meanings of the existence 
of a joint distribution $\rho(\tilde{\lambda})$ 
which returns $\rho_{pq}(\lambda,\lambda_p,\lambda_q)$
as marginal distributions.
The existence of such a joint distribution $\rho(\tilde{\lambda})$ 
is equivalent to the existence of deterministic local realistic
models which reproduce 
$\rho_{pq}(\lambda,\lambda_p,\lambda_q)$$^{(13)}$.
That is, 
 if $\rho_{pq}(\lambda,\lambda_p,\lambda_q)$ 
can be obtained from a single joint distribution as marginals then these
$\rho_{pq}(\lambda,\lambda_p,\lambda_q)$ can be reproduced
by some local realistic models, and
 if $\rho_{pq}(\lambda,\lambda_p,\lambda_q)$ cannot be obtained
 from a single joint distribution as marginals then these
$\rho_{pq}(\lambda,\lambda_p,\lambda_q)$ cannot be reproduced
by any local realistic models. 
Thus we may call $\rho_{pq}(\lambda,\lambda_p,\lambda_q)$
which have such a joint distribution {\it local correlations}
 of $\lambda,\lambda_p$ and $\lambda_q$, and call
$\rho_{pq}(\lambda,\lambda_p,\lambda_q)$ which does not have
such a joint distribution {\it nonlocal correlations} of 
$\lambda, \lambda_p$ and $\lambda_q$. 
On the other hand, we could show that Bell inequality can be recovered
(Eq.(\ref{20}) ) in the case of these local correlations
of $\lambda, \lambda_p$ and $\lambda_q$, while we could not
show that Bell inequality can be recovered in the case of 
nonlocal correlations
of $\lambda, \lambda_p$ and $\lambda_q$.
This fact is in accord with the fact that Bell inequality
is obtained for {\it local} realistic models.

\section{Discussion and conclusion}
It is obvious that Bell inequality cannot be recovered without
some constraints on $\rho_{pq}(\lambda,\lambda_p,\lambda_q)$
( $ p=a,a^\prime \hspace{0.3cm} q=b,b^\prime $ )
from locality. In previous section, we could show that Bell inequality
can be recovered in two cases.
(i)  $\rho_{pq}(\lambda,\lambda_p,\lambda_q)$
  are factorized (Eq.(\ref{14})),
(ii) there exists a joint distribution $\rho(\tilde{\lambda})$
which returns 
 $\rho_{pq}(\lambda,\lambda_p,\lambda_q)$
 as marginal distributions (Eq.(\ref{16})).
 By the way, case(i) is a subset of case(ii),
 because Eq.(\ref{14}) can be reproduced by 
 \begin{equation}
   \rho(\tilde{\lambda})=
    \rho(\lambda)
    \rho_a(\lambda_a)\rho_{a^\prime}(\lambda_{a^\prime})
    \rho_b(\lambda_b)\rho_{b^\prime}(\lambda_{b^\prime}).
    \label{21}
 \end{equation}
In the case of (i), the distribution of $\lambda_p$ and $\lambda_q$
are independent each other. In the case of (ii), the distribution
of $\lambda_p$ and $\lambda_q$ may have correlations, unless they are 
nonlocal one.
By the way, in the Bell's later work$^{(7)}$  the independency
case ( condition (i)) is considered. In this connection, we generalized
the Bell's later work to the case of condition (ii).

Summing up the above results, we could say that if we ignore $\lambda_i$
( $i=a,a^\prime,b,b^\prime$ ) the three following ones are reduced to 
an identical one. (a) stochastic models without $\lambda_i$$^{(15,13,14)}$ 
(b) deterministic models with $\lambda_i$
( which are correlated ( condition (i) ) or not ( condition (ii) ).
(c) stochastic models with $\lambda_i$$^{(7)}$.

 With nonlocal correlations 
 $\rho_{pq}(\lambda,\lambda_p,\lambda_q)$
 of $\lambda$ and $\lambda_i$
, Bell inequality was not obtained.
In other words, if the distributions of $\lambda$ and
$\lambda_i$ are nonlocal ( or do not
have a joint distribution ) then they can give outcomes that violate
the Bell inequality (Eq.(\ref{9})) for which $\lambda_i$ acted as measuring
equipments.
Recently, it was shown
that violation of the Bell inequality does not necessarily mean 
that all the source particle pairs are nonlocal ones$^{(16)}$,
that is, the nonlocality can be ascribed to some subsets of ensemble of 
particle pairs.  
Here we have shown that the nonlocality might be ascribed to 
spin-measuring equipments.

In summary, the original supplementary variable $\lambda$ is not to
be regarded to include supplementary variables in spin-measuring
equipments as shown in section 3. Thus supplementary
variables should be introduced additionally in spin-measuring equipments. 
When the supplementary variables 
introduced in spin-measuring equipments have local correlations, the
Bell inequality is recovered. On the other hand, when they have nonlocal
correlations, the Bell inequality is not recovered. This fact is in 
accord with the fact that the Bell inequality is derived for local 
realistic models.

\end{document}